\documentclass[12pt]{article}
\usepackage{graphicx}

\def\Title#1{\begin{center} {\Large #1 } \end{center}}
\def\Author#1{\begin{center}{ \sc #1} \end{center}}
\def\Address#1{\begin{center}{ \it #1} \end{center}}

\newcommand\pubblock{\rightline{\begin{tabular}{l}   \end{tabular}}}
\newcommand{\mev}{\ensuremath{\mathrm{\,Me\kern -0.1em V}}\xspace}

\newenvironment{Abstract}{\begin{quotation}  }{\end{quotation}}

\newenvironment{Presented}{\begin{quotation} \begin{center} 
             PRESENTED AT\end{center}\bigskip 
      \begin{center}\begin{large}}{\end{large}\end{center} \end{quotation}}

\def\kaon  {\ensuremath{K}\xspace}
\def\Kbar  {\kern 0.2em\overline{\kern -0.2em K}{}\xspace}

\def\Kz    {\ensuremath{\kaon^0}\xspace}
\def\Kzb   {\ensuremath{\Kbar^0}\xspace}
\def\KzKzb {\ensuremath{\Kz \kern -0.16em \Kzb}\xspace}
\def\Kp    {\ensuremath{\kaon^+}\xspace}
\def\Km    {\ensuremath{\kaon^-}\xspace}

\def\KpKm  {\ensuremath{\Kp \kern -0.16em \Km}\xspace}

\def\B       {\ensuremath{B}\xspace}
\def\Bbar    {\ensuremath{\kern 0.18em\overline{\kern -0.18em \B}{}}\xspace}

\def\Lbar {\ensuremath{\kern 0.1em\overline{\kern -0.1em\Lambda}}\xspace}

\def\beq{\begin{equation}}
\def\eeq#1{\label{#1}\end{equation}}
\def\eeqn{\end{equation}}

\def\beqa{\begin{eqnarray}}
\def\eeqa#1{\label{#1}\end{eqnarray}}
\def\eeqan{\end{eqnarray}}

\let\bar=\overbar

\def\Dslash{\not{\hbox{\kern-4pt $D$}}}
\def\dslash{\not{\hbox{\kern-2pt $\del$}}}

\def\msb{{\bar{\ssstyle M \kern -1pt S}}}

\begin{document}
\begin{titlepage}
\pubblock

\vfill
\Title{Charmless B decays in modes with similar tree and penguin contributions }
\vfill
\Author{ Ignacio Bediaga, on the  behalf of the  LHCb Collaboration. }
\Address{ Centro Brasileiro de Pesquisas F\'isicas, Rio de Janeiro, Brazil}
\vfill
\begin{Abstract}

Charmless $B$ decays are dominated by contributions from the short
distance amplitudes  from tree level and penguin loop-level
amplitudes. The Tree contribution  presents  a weak phase $\gamma$.
The interference  between these two amplitudes can generate a CP
asymmetry depending on the relative contribution of each amplitude. 
In multi-body charmless $B$ decays, these relative contributions
can change along the phase space,  given a non isotropic distribution
of   CP asymmetries in the Dalitz plot.  Two recent LHCb analyses
involving charmless multi-body B decays are discussed: the obsevation
of CP asymmetries in the phase space  of the three-body decays  $B^\pm
\to \pi^\pm \pi^+ \pi^-$ and  $B^\pm \to \pi^\pm K^+ K^-$; and the angular analysis of the  
 $B^0 \to \phi K^*(892)^0$ decay.

\end{Abstract}
\vfill
\begin{Presented}
FPCP-2014- Flavor Physics \& CP Violation\\
Marseille, France,  May  25-30, 2014
\end{Presented}
\vfill
\end{titlepage}
\def\thefootnote{\fnsymbol{footnote}}
\setcounter{footnote}{0}

\section{Introduction}

Multi-body hadronic $B$ decays have been showing some interesting
physics signatures in both; flavor physics and CP violation. The large
phase space of charmless $B$ decays together with LHCb high
statistics, allows  observations of  some important distributions
related to hadrons interactions. The  polarization amplitudes to the $
B^0 \to \phi K^*(892)^0$ decay \cite{AngleLHCb}
 and the high   CP violation observed  inside the phase space of the
 decays $B^\pm \to \pi^\pm \pi^+ \pi^-$ and  $B^\pm \to \pi^\pm K^+
 K^-$ \cite{KKpiand3pi} are examples of these new signatures. 

The $ B^0 \to \phi K^*(892)^0$ decay involves a spin-0 B-meson decaying into two spin-1
vector mesons. Due to angular momentum conservation there are only three independent 
configurations of the final-state spin vectors, a longitudinal
component where in the $B^0$ rest frame both resonances are polarized in their
direction of motion, and two transverse components with collinear and 
orthogonal polarizations. Other than measurements of the polarization amplitudes, 
triple-product asymmetries were studied by LHCb Collaboration \cite{AngleLHCb}.

 Angular analyses have shown that the longitudinal and transverse components in this decay
have roughly equal amplitudes. Similar results are seen in other $B\to VV$ 
penguin transitions
\cite{delAmoSanchez:2010mz,Abe:2004mq, Aubert:2006fs,LHCb-PAPER-2011-012}. 
The different behavior of tree and penguin decays has attracted much
theoretical attention, with several explanations    
proposed such as large contributions from penguin annihilation effects \cite{kagan} or final-state
interactions \cite{PhysRevD.76.034015}. More recent
calculations based on QCD factorization \cite{beneke,PhysRevD.80.114026} are  
consistent with the data, although with significant uncertainties. 

Charmless decays of $B$ mesons to three hadrons are dominated by quasi-two-body processes involving intermediate resonant states. 
The rich interference pattern present in such decays makes them favorable for the investigation of  CP asymmetries that are localized in the phase space~\cite{Miranda1,Miranda2}.
The large samples of charmless $B$ decays collected by the LHCb experiment allow direct CP violation to be measured in regions of phase space.   
In previous measurements of this type, the phase spaces of  $B^\pm \to K^\pm \pi^+ \pi^-$ and  $B^\pm \to K^\pm K^+ K^-$  decays were observed to have regions of large local asymmetries~\cite{KKpiand3pi,LHCb-PAPER-2013-027}. 

Recent efforts have been made to understand the origin of the large asymmetries. 
For direct CP violation to occur, two interfering amplitudes with different weak and strong phases must be involved in the decay process~\cite{BSS1979}.  
Interference between intermediate states of the decay can introduce large strong phase differences, and is one mechanism for explaining local asymmetries in the phase space~\cite{PhysRevD.87.076007,Bhattacharya:2013cvn}.
Another explanation focuses on final-state $KK \leftrightarrow \pi \pi$ rescattering, which can occur between decay channels with the same flavor quantum numbers~\cite{LHCb-PAPER-2013-027,Bhattacharya:2013cvn,IgnacioCPT}. 
Invariance of CPT symmetry constrains hadron rescattering so that the sum of the partial decay widths, for all channels with the same final-state quantum numbers related by the S matrix, must be equal for charge-conjugated decays.  
Effects of SU(3) flavor symmetry breaking have also been investigated and partially explain the observed patterns~\cite{Bhattacharya:2013cvn,Xu:2013dta,Gronau:2013mda}.

The studies reported here are performed using $pp$ collision data, corresponding to an integrated 
luminosity of $1.0~fb^{-1}$, collected at a centre-of-mass energy of 
$7$ MeV  with the LHCb detector.

\section{\boldmath $ B^0 \to \phi K^*(892)^0$  signal and the angular fit model}

The selection of starts from well reconstructed charged particles with a $p_{\rm T} > 500 {\rm ~ MeV}$
that traverse the entire spectrometer~\cite{LHCb-PAPER-2013-010}. Fake tracks, not associated to actual charged particles, are suppressed using the output 
of a neural network trained to discriminate between these and real particles. To select well-identified 
pions and kaons, the difference in the logarithms of the 
likelihood of the kaon hypothesis relative to the pion
hypothesis is used. The resulting charged particles are combined to form $\phi$
and $K^*(892)^0$ meson candidates. The invariant mass of the $K^+K^-$
($K^+\pi^-) $ pair is required to be within $\pm 15 {\rm ~MeV}/c^2 $ ($\pm 150
{\rm ~ MeV}/c^2$) of the known mass of the  $\phi$ ( $K^*(892)^0$) meson \cite{PDG2012}. 
Finally, the $p_{T}$ of the $\phi$ and $K^*(892)^0$ mesons should both be greater than $900 {\rm MeV}$, 
and the fit of their two-track vertices should have a $\chi^2<9$.  

\begin{figure}[htb]
\centering
\includegraphics[height=2in]{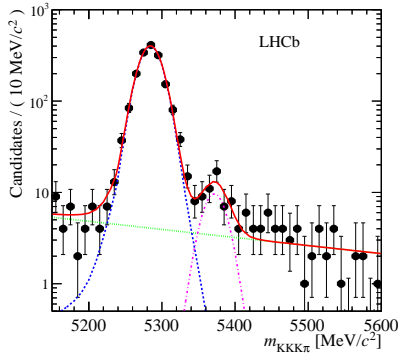}
\caption{\small Invariant mass distribution for selected $K^+K^-K^+\pi^-$ 
  candidates. A fit to the model described in the text is superimposed 
  (red solid line). The signal contribution is shown as the blue
  dotted line. The contribution from combinatorial background is shown in green 
  (dotted line). A contribution from $ B^0_s \to \phi K^*(892)^0$(purple dot-dashed line) decays 
  is visible around the known $B_s^0$ meson mass.}
\label{fig:bdmass}
\end{figure}

The signal yield is determined from  an unbinned maximum likelihood fit to the $K^+K^-K^+\pi^-$ invariant mass distribution.
The selected mass range is chosen to avoid modeling partially reconstructed $B$ decays with a missing hadron or photon.
In the fit the signal invariant mass distribution is modeled as the sum of
a Crystal Ball function \cite{Skwarnicki:1986xj} and a wider Gaussian function with a common mean. 
The width and fraction of the Gaussian function are fixed to values obtained using
simulated events. A component is also included to account for
the small contribution from the decay $ B^0_s \to \phi K^*(892)^0$   
\cite{LHCb-PAPER-2013-012}. The shape parameters for this
component are in common with the $B^0$ signal shape and the relative position of the $B^0_s$ signal 
with respect to the $B^0$ signal is fixed using the known mass difference between $B^0_s$ and $B^0$ mesons \cite{PDG2012}. The invariant mass distribution is shown in Fig.~\ref{fig:bdmass},
together with the result of the fit, from which a yield of $1655 \pm
42$ $B^0$ signal candidates is found. The background is mainly combinatorial and is modeled by an exponential. Possibles background from
$B^0\rightarrow \phi \phi$ decays,  $B^0\rightarrow D_s^+ K^- (D_s^+\to \phi\pi^+)$ decays, 
which would peak in the signal region, is found to be negligible.

An angular analysis is performed with  the $ B^0 \to \phi K^*(892)^0$ signal events obtained with the  selection above, we performed an angular analysis.  The physics parameters of 
interest for this analysis are defined in Table~\ref{tab:parameters}. They include the polarization amplitudes, 
phases and amplitude differences between $B^0$ and $\bar B^0$  decays. 

The amplitudes and phases can be used to calculate triple-product
asymmetries~\cite{AngleLHCb}. Non-zero
triple-product asymmetries arise either due to a $T$-violating phase or a CP-conserving phase
and final-state interactions. Assuming CPT symmetry, a $T$-violating phase,
which is a \textit{true} asymmetry, implies that CP is violated. Otherwise \textit{fake} asymmetry indicate CP conservation and the presence of final state interaction \cite{gronau}.

The acceptance of the detector is not uniform as a function of
the decay angle of the $K^+\pi^-$ system~($\theta_1$) and 
the $K^+\pi^-$ invariant mass. This   acceptance is modelled using a four-dimensional function that depends on the 
three decay angles and the $K^+\pi^-$ invariant mass. 
The shape of this function is obtained from simulated data.

\section{Results}

The fit function incorporate other than the P-wave contribution,   for the first time the S-wave component is include in this kind of study. 
The differential decay width depends on the invariant masses of
the $K^+\pi^-$ and $K^+K^-$ systems, denoted $m_{K\pi}$ and $m_{KK}$, respectively. The P-wave $K^+\pi^-$ amplitude is parameterized using a
relativistic spin-$1$ Breit-Wigner resonance function. The P-wave $K^+K^-$ amplitude is modeled
in a similar way using the values $m_0(\phi) = 1019.455 {\rm ~ MeV}$
 and $\Gamma_0(\phi) = 4.26 {\rm ~ MeV}$~\cite{PDG2012}. The S-wave contributions in both the $K^+K^-$
and $K^+\pi^-$ system are considered. The default fit uses the LASS parameterization to model the $K^+\pi^-$
S-wave \cite{LASS}. As variations of this, both a pure phase-space model
and a spin-0 relativistic Breit-Wigner with mean and width of the
$K^*_0(1430)$ resonance are considered~{\cite{PDG2012}}. For the $K^+K^-$ S-wave
a pure phase-space model is tried in place of the Flatt\'e
parameterization  \cite{flatte}. The largest observed deviation from the
nominal fit is taken as a systematic uncertainty \cite{AngleLHCb}.

\begin{figure}[htb!]
\begin{center}
\includegraphics[width = 0.48\columnwidth]{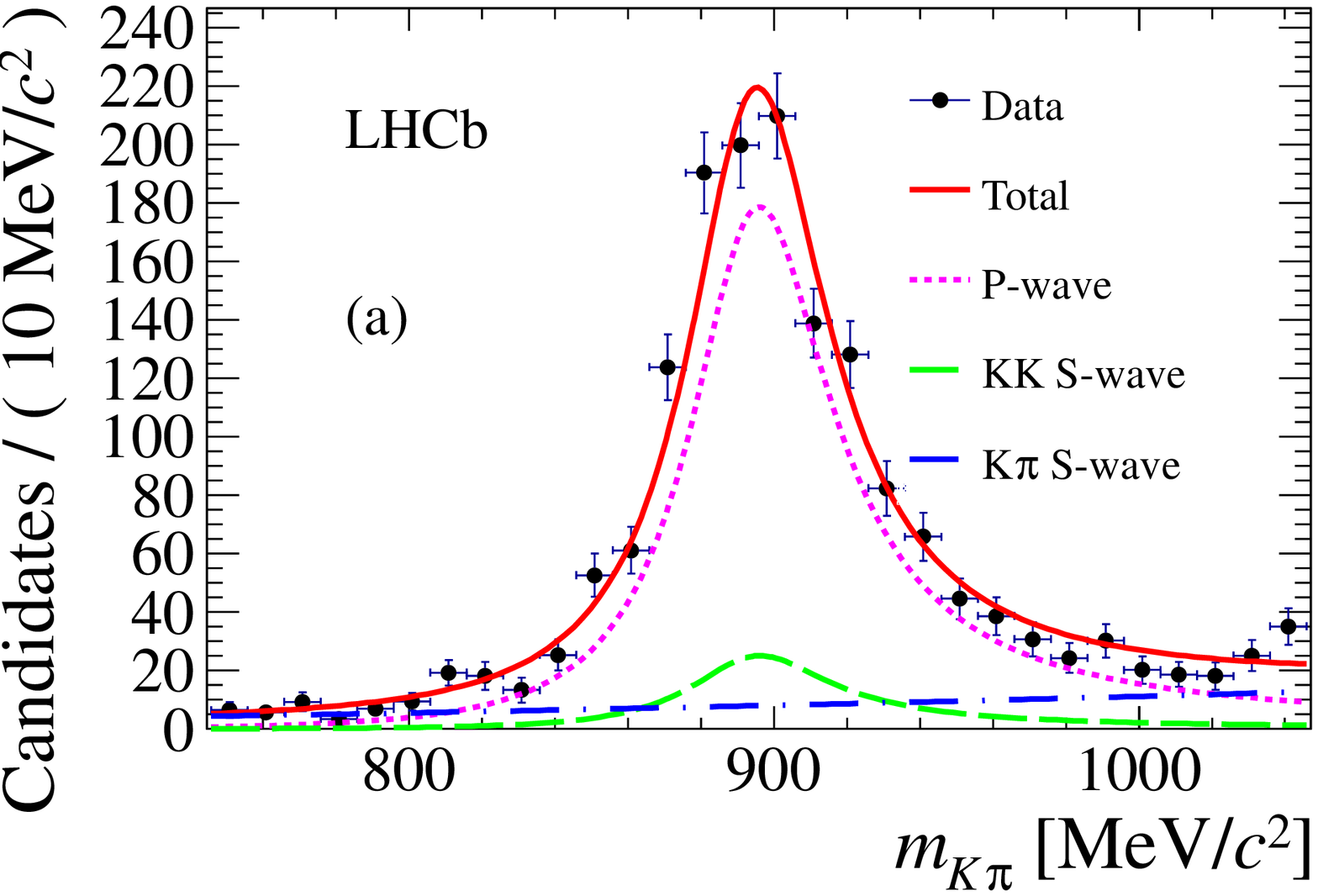}
\includegraphics[width = 0.48\columnwidth]{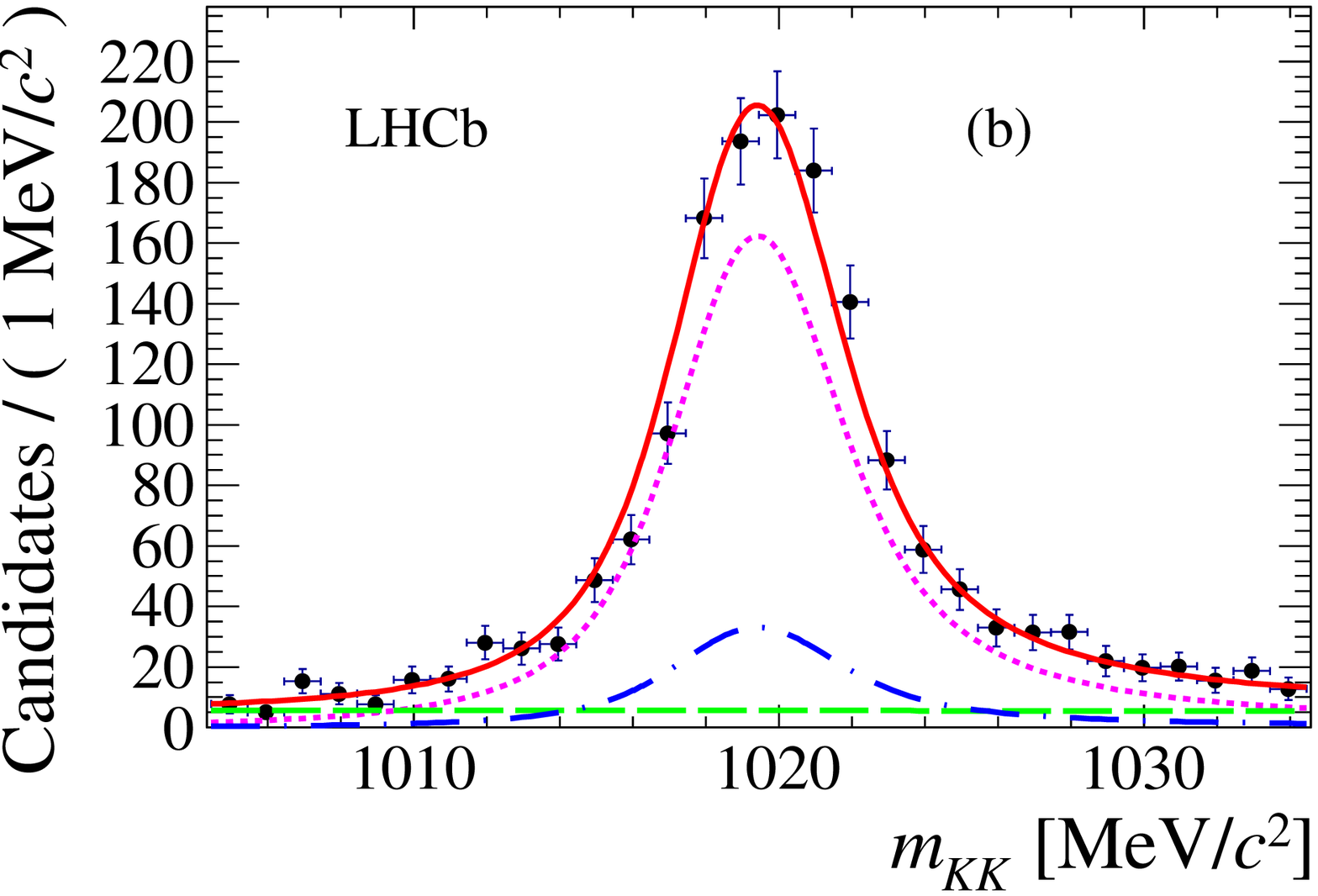}
\includegraphics[width = 0.48\columnwidth]{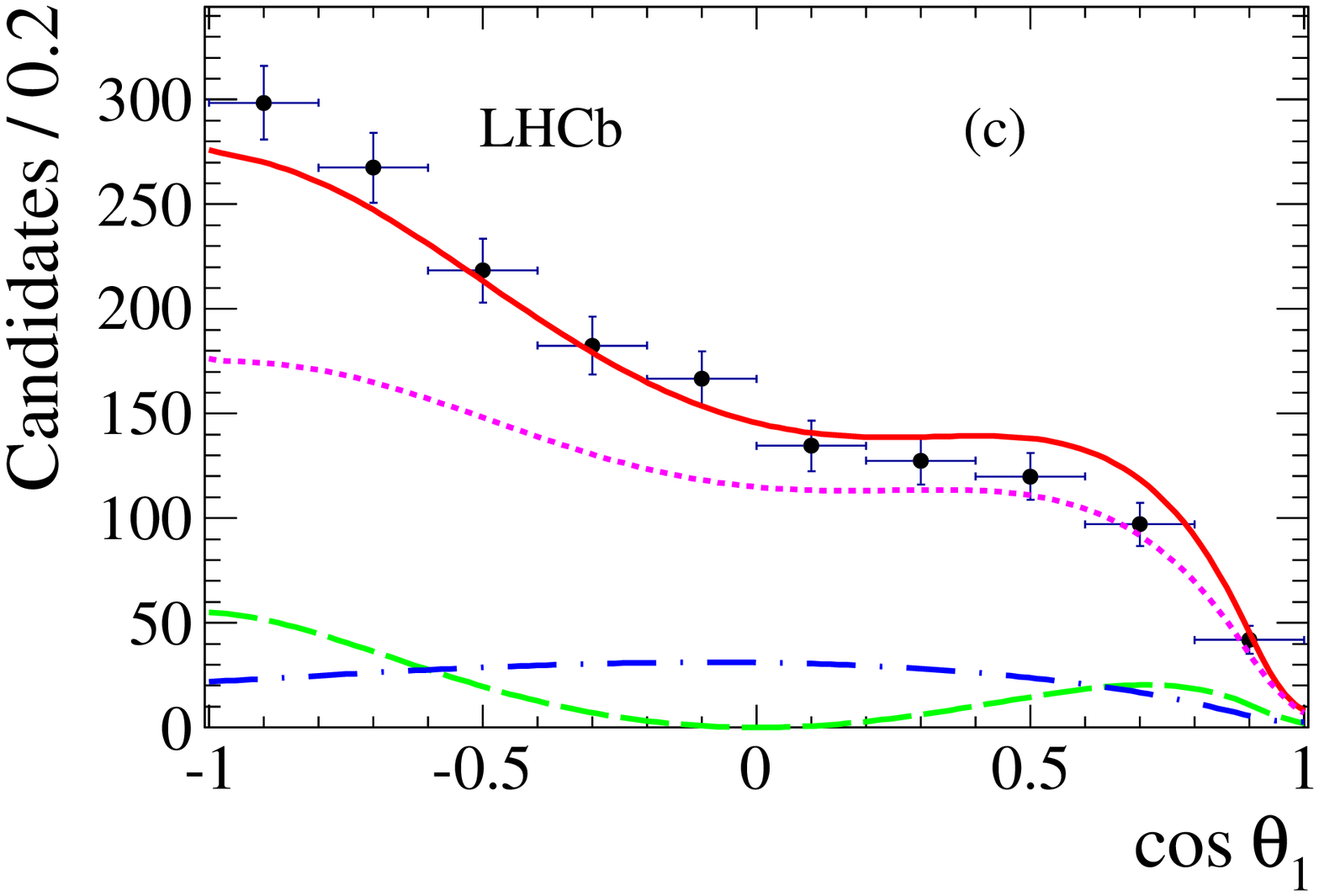}
\includegraphics[width = 0.48\columnwidth]{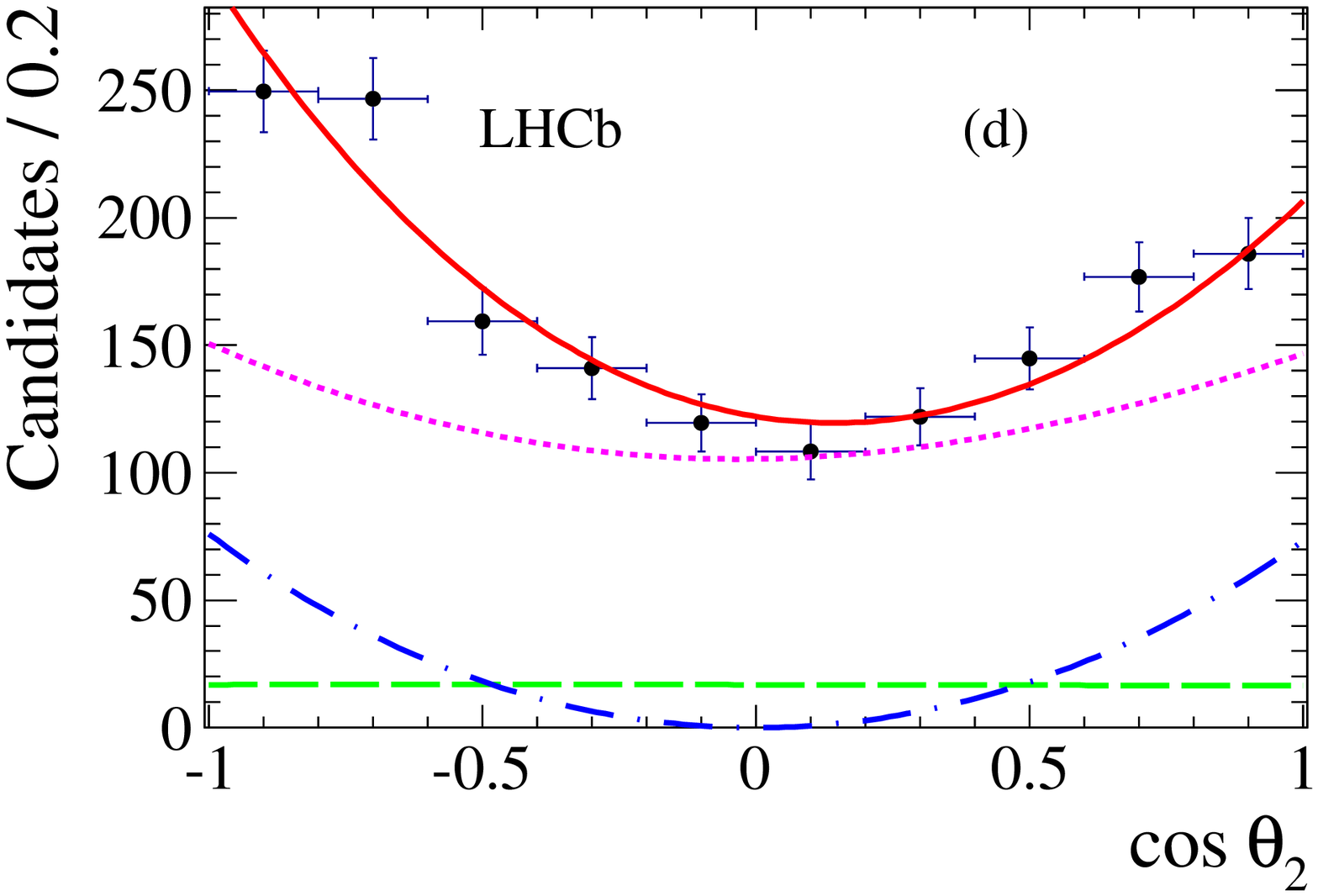}
\includegraphics[width = 0.48\columnwidth]{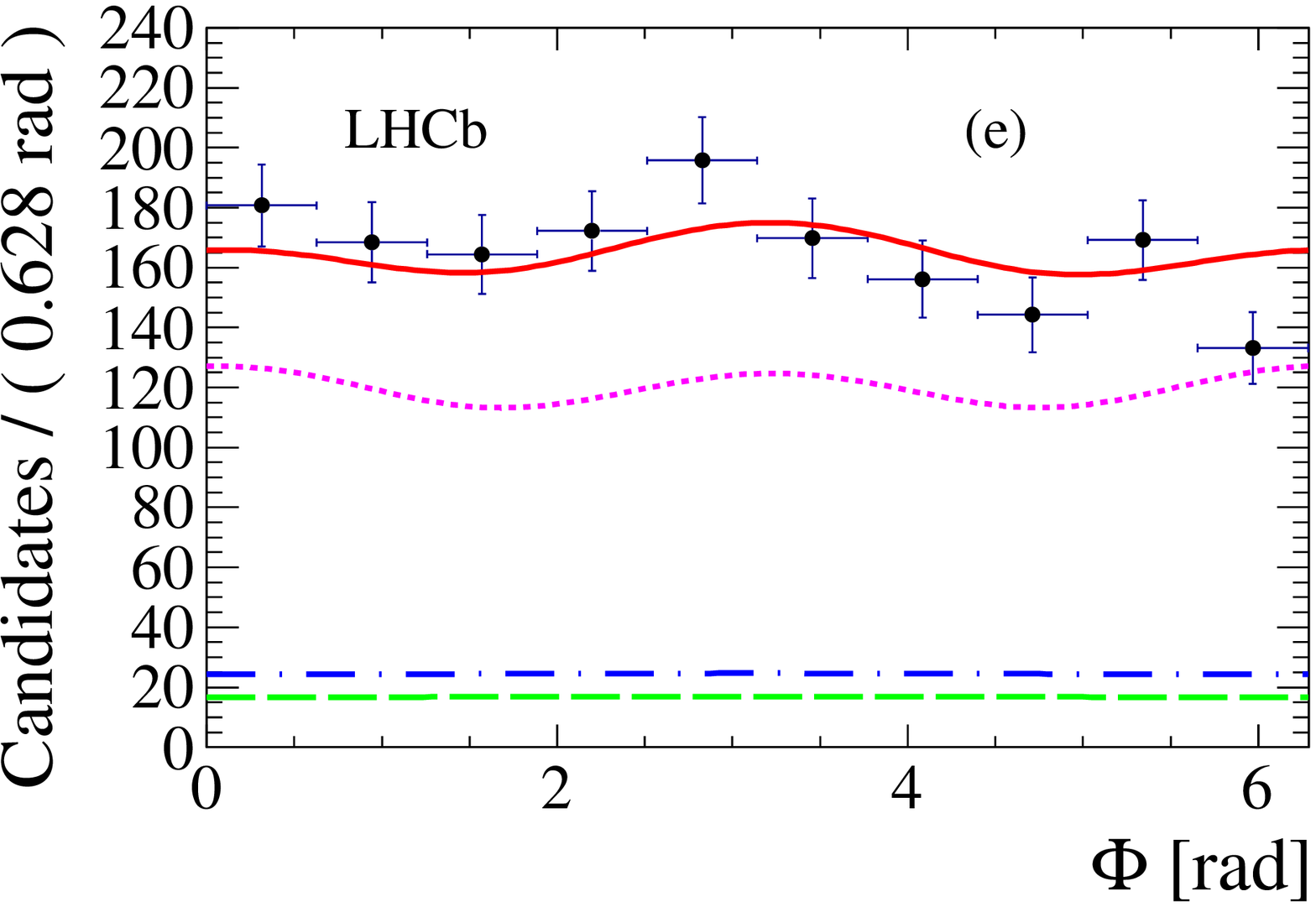}
\caption{\small Data distributions for the helicity angles and of the
intermediate resonance masses:  (a) $m_{K\pi}$ and (b) $m_{KK}$, (c) $\cos \theta_1$, (d) $\cos\theta_2$ and 
(e) $\Phi$. The background has been 
subtracted using the SPlot technique. The results of the fit are superimposed.
\label{fig:fitproj}}
\end{center}
\end{figure}

Figure \ref{fig:fitproj} shows the data distributions for the intermediate resonance masses and helicity angles with the projections of the best fit
overlaid. The goodness of fit is estimated using a point-to-point dissimilarity
test \cite{Williams:2010vh}, the corresponding $p$-value is 0.64. 

The fit results are listed in Table~\ref{tab:parameters}. The longitudinal distribution $f_L$, defined in \cite{AngleLHCb},   returned by the fit is close to 0.5, indicating that the
longitudinal and transverse polarizations ($f_{\perp}$), have similar size. Significant S-wave 
contributions are found in both the $K^+\pi^-$ and $K^+K^-$
systems. The CP  asymmetries in both the amplitudes and 
the phases are consistent with zero.

The largest systematic uncertainties on the results of the angular analysis 
arise from the understanding of the detector acceptance. 
The angular acceptance function is determined from simulated
events. This and other  systematic uncertainties are discussed in Ref. \cite{AngleLHCb}.

\begin{table}[!htb]
\begin{center}
\caption{\small Angular analysis results. The  parameters are defined in Ref. \cite{AngleLHCb} . 
The first and second uncertainties are statistical and systematic, respectively. 
\label{tab:parameters}}
\begin{tabular}{ccc}
\hline
Parameter & Definition & Fitted value \\ 
\hline 
\\ [-0.48cm]
$f_\textrm{L}$ & $0.5(|A_0|^2/F_\textrm{P} + |\overline{A}_0|^2/\overline{F}_\textrm{P})$ &\phantom{-} 
$0.497\pm 0.019\pm 0.015$ \\
$f_{\perp}$ & $0.5(|A_\perp|^2/F_\textrm{P} + |\overline{A}_\perp|^2/\overline{F}_\textrm{P})$ & \phantom{-}
$0.221\pm 0.016\pm 0.013$ \\
$f_{\textrm{S}}(K\pi)$  & $0.5(|A_{\textrm{S}}^{K\pi}|^2 + |\overline{A}_{\textrm{S}}^{K\pi}|^2)$ & \phantom{-}
$0.143\pm 0.013\pm 0.012$\\
$f_{\textrm{S}}(KK)$  & $0.5(|A_{\textrm{S}}^{KK}|^2 + |\overline{A}_{\textrm{S}}^{KK}|^2)$ & \phantom{-}
$0.122\pm 0.013\pm 0.008$ \\
\\ [-0.48cm]
$\delta_\perp$     & $0.5(\arg{A_\perp} + \arg{\overline{A}_\perp})$ & \phantom{-}
$2.633\pm 0.062\pm 0.037$ \\
$\delta_\parallel$     & $0.5(\arg{A_\parallel} + \arg{\overline{A}_\parallel})$ &\phantom{-} 
$2.562\pm 0.069\pm 0.040$ \\
$\delta_{\textrm{S}}(K\pi)$     & $0.5(\arg{A_{\textrm{S}}^{K\pi}} + \arg{\overline{A}_{\textrm{S}}^{K\pi}})$ & \phantom{-}
$2.222\pm 0.063\pm 0.081$ \\
$\delta_{\textrm{S}}(KK)$     & $0.5(\arg{A_{\textrm{S}}^{KK}} + \arg{\overline{A}_{\textrm{S}}^{KK}})$ & \phantom{-}
$2.481\pm 0.072\pm 0.048$\\ 
\\ [-0.48cm]
$\mathcal{A}_{0}^{CP}$         & $(|A_0|^2{/F_\textrm{P}} -
|\overline{A}_0|^2{/\overline{F}_\textrm{P}})/(|A_0|^2{/F_\textrm{P}}
+ |\overline{A}_0|^2{/\overline{F}_\textrm{P}})$ & 
$-0.003\pm 0.038\pm 0.005$ \\
$\mathcal{A}_{\perp}^{CP}$         & $(|A_\perp|^2{/F_\textrm{P}} -
|\overline{A}_\perp|^2{/\overline{F}_\textrm{P}})/(|A_\perp|^2{/F_\textrm{P}}
+ |\overline{A}_\perp|^2{/\overline{F}_\textrm{P}})$ &
$+0.047\pm 0.074\pm 0.009$ \\
$\mathcal{A}_{S}(K\pi)^{CP}$ & $(|A_{\textrm{S}}^{K\pi}|^2 - |\overline{A}_{\textrm{S}}^{K\pi}|^2)/(|A_{\textrm{S}}^{K\pi}|^2 + |\overline{A}_{\textrm{S}}^{K\pi}|^2)$ &
$+0.073\pm 0.091\pm 0.035$ \\
$\mathcal{A}_{S}(KK)^{CP}$   & $(|A_{\textrm{S}}^{KK}|^2 - |\overline{A}_{\textrm{S}}^{KK}|^2)/(|A_{\textrm{S}}^{KK}|^2 + |\overline{A}_{\textrm{S}}^{KK}|^2)$ & 
$-0.209\pm 0.105\pm 0.012$ \\
\\ [-0.48cm]
$\delta_\perp^{CP}$          & $0.5(\arg{A_\perp} - \arg{\overline{A}_\perp})$ & 
$+0.062\pm 0.062\pm 0.005$ \\
$\delta_\parallel^{CP}$      & $0.5(\arg{A_\parallel} - \arg{\overline{A}_\parallel})$ & 
$+0.045\pm 0.069\pm 0.015$ \\
$\delta_{S}(K\pi)^{CP}$      & $0.5(\arg{A_{\textrm{S}}^{K\pi}} - \arg{\overline{A}_{\textrm{S}}^{K\pi}})$ & 
$+0.062\pm 0.062\pm 0.022$ \\
$\delta_{S}(KK)^{CP}$        & $0.5(\arg{A_{\textrm{S}}^{KK}} - \arg{\overline{A}_{\textrm{S}}^{KK}})$ &
$+0.022\pm 0.072\pm 0.004$ \\\hline 
\end{tabular}
\end{center}
\end{table}

The results for the P-wave parameters are consistent with, but more precise
than previous measurements. 
All measurements are consistent with the presence of a large transverse component 
rather than the naive expectation of a dominant longitudinal polarization. 
It is more difficult to make comparisons for the S-wave components as this 
is the first measurement to include consistently the effect of the S-wave in the $K^+K^-$ system, and because 
the $K^+\pi^-$ mass range is different with respect to the range used in previous analyses.

The  values for the triple-product asymmetries  are derived from the measured parameters 
and given in Table~\ref{tab:tripleproducts}. 
The true asymmetries are consistent with zero, showing no evidence for physics beyond the Standard Model.  In contrast, all but one of the fake asymmetries are significantly different from zero, 
indicating the presence of final-state interactions.

\begin{table}[!htb]
\begin{center}
\caption{\small Triple-product asymmetries. The first and second uncertainties on the measured values 
are statistical and systematic, respectively. \label{tab:tripleproducts}}
\begin{tabular}{cc}
\hline
Asymmetry & Measured value \\ 
\hline 
$A_{T}^1$(true) & $-0.007\pm 0.012\pm 0.002$ \\
$A_{T}^2$(true) & $+0.004\pm 0.014\pm 0.002$ \\
$A_{T}^3$(true) & $+0.004\pm 0.006\pm 0.001$ \\
$A_{T}^4$(true) & $+0.002\pm 0.006\pm 0.001$ \\ 
$A_{T}^1$(fake) & $-0.105\pm 0.012\pm 0.006$ \\
$A_{T}^2$(fake) & $-0.017\pm 0.014\pm 0.003$ \\
$A_{T}^3$(fake) & $-0.063\pm 0.006\pm 0.005$ \\
$A_{T}^4$(fake) & $-0.019\pm 0.006\pm 0.007$ \\  \hline
\end{tabular}
\end{center}
\end{table}

\section{ \boldmath $ B^\pm \to \pi^\pm K^+  K^- $ and $ B^\pm \to \pi^\pm \pi^+  \pi^- $   inclusive CP asymmetries }

The inclusive CP asymmetries involves the difference of total number of events of the negative  and negative $B^\pm$ three body decays divided by the sum of them. For these decays are required to satisfy a set of selection criteria related with  their transverse momenta, vertex and tracking quality. 
Final-state kaons and pions are further selected using particle identification information, provided by two ring-imaging Cherenkov detectors~\cite{LHCb-DP-2012-003}, and are required to be incompatible with a muon~\cite{LHCb-DP-2013-001}. 
The kinematic selection is  common to both decay channels, while the
particle identification selection are  specific to each final
state. Charm contributions are removed by excluding the regions  of
$\pm 30 {\rm ~ MeV}$ around the world average value of the $D^0$ mass~\cite{PDG2012} in the two-body invariant masses $m_{\pi\pi}$, $m_{K\pi}$ and $m_{KK}$. 

\begin{figure*}[tb]
\centering
\includegraphics[width=0.48\linewidth]{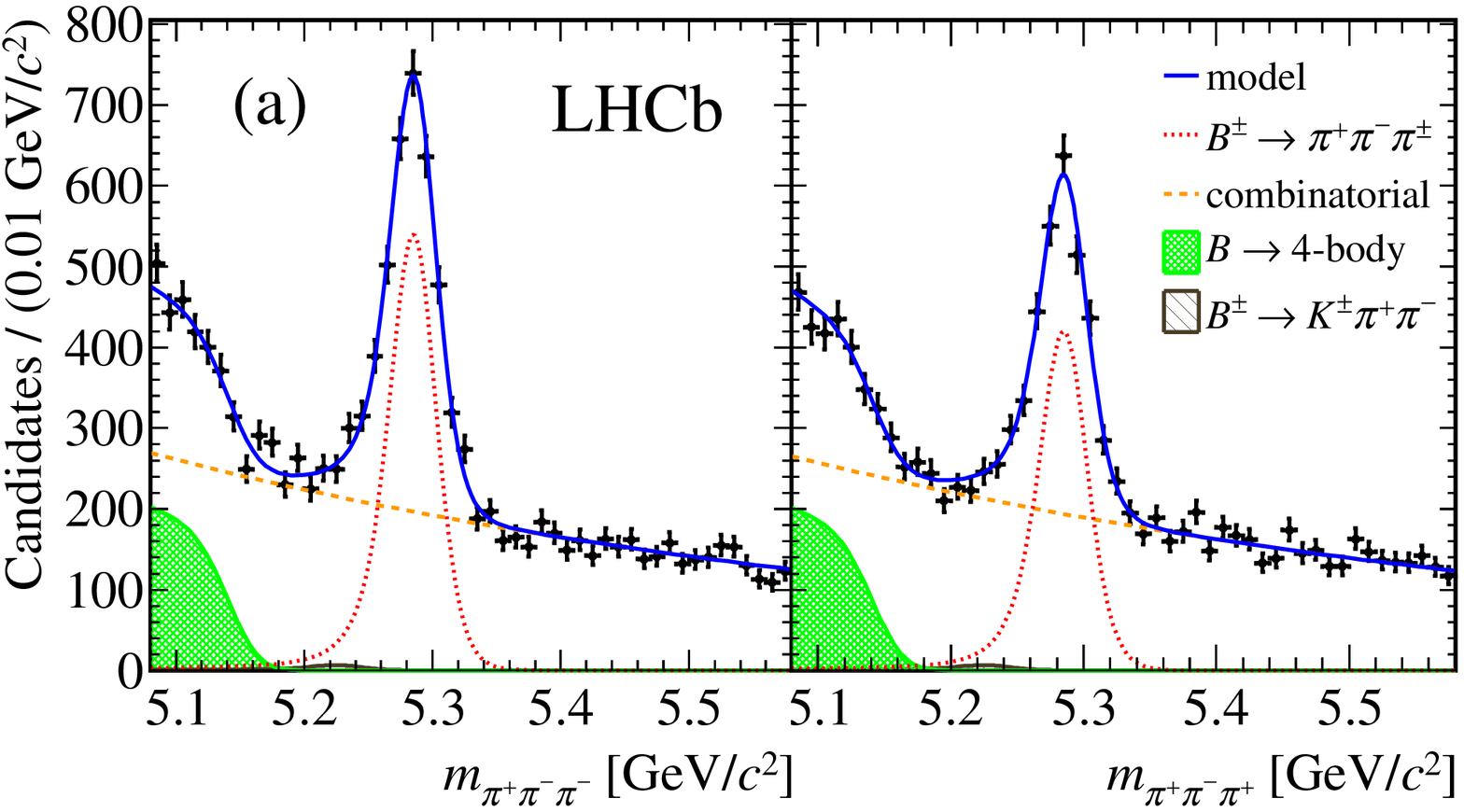}
\hspace{0.2cm}
\includegraphics[width=0.48\linewidth]{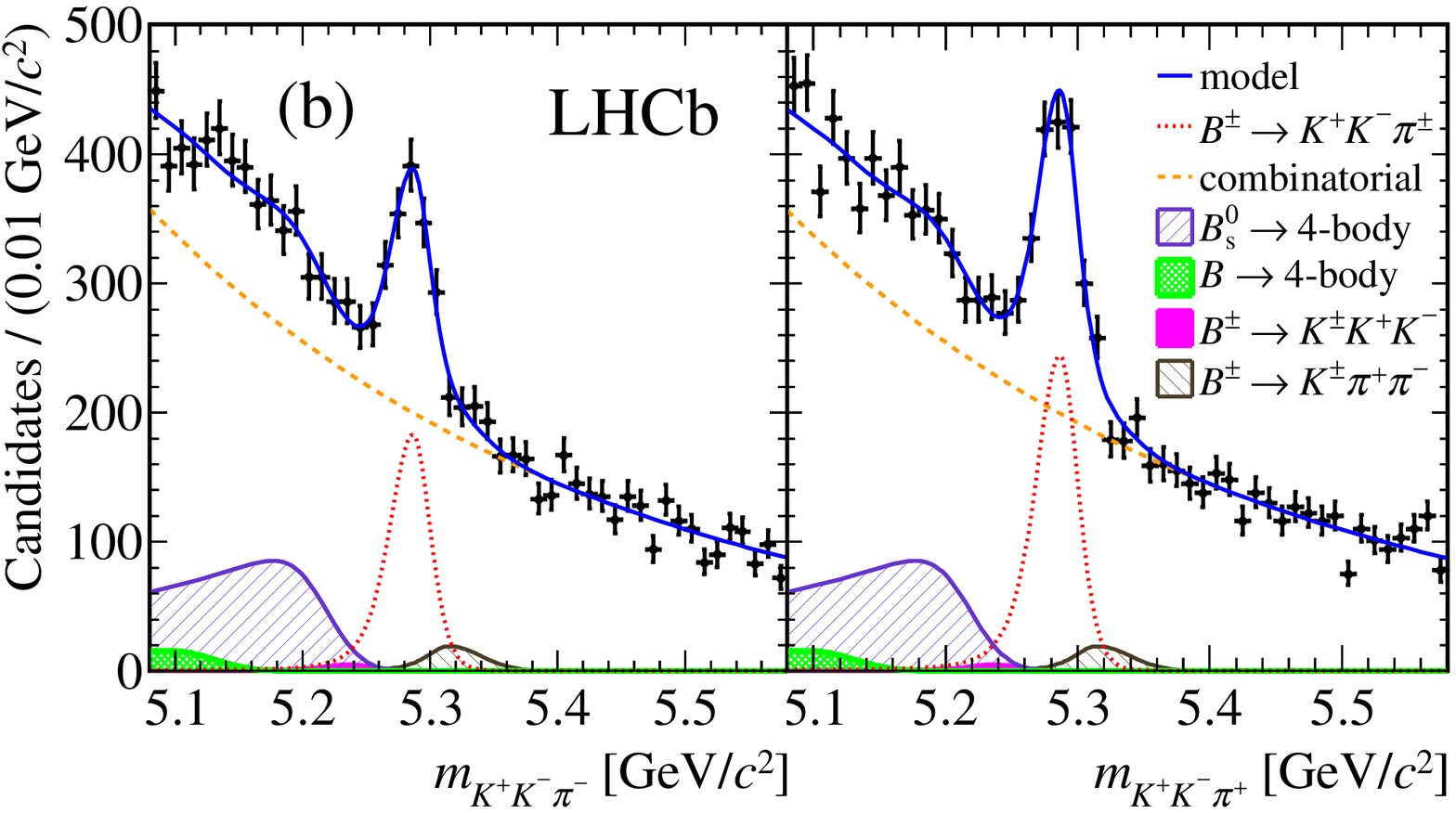}
\caption{Invariant mass spectra of (a) $\pi^\pm \pi^+ \pi^- $  decays and (b) $\pi^\pm K^+  K^- $ decays. The left panel in each figure shows 
the $B^-$ modes and the right panel shows the $B^+$ modes.
The results of the unbinned maximum likelihood fits are overlaid. The main components of the fit are also shown. 
}
\label{MassFit}
\end{figure*}

Unbinned extended maximum likelihood fits to the mass spectra of the selected $B^\pm$ candidates are performed to obtain the signal yields and raw asymmetries. 
Both  signals components are parametrized by a {\mbox{Cruijff}} function~\cite{Cruijff} with equal left and right widths and different radiative tails to account for the asymmetric effect of final-state radiation on the signal shape. 
The means and widths are allowed to vary freely in the fit, while the tail parameters are fixed to the values obtained from simulation. 
The combinatorial background is described by an exponential distribution whose parameter is left free in the fit.
The backgrounds due to partially reconstructed four-body $B$ decays are parametrized by an ARGUS distribution~\cite{Argus} convolved with a Gaussian resolution function.
For the $\pi^\pm \pi^+ \pi^-$  modes the shape and yield parameters describing the backgrounds are varied in the fit, while for  $\pi^\pm K^+  K^- $ decays they are taken from simulation, due to a further contribution from four-body $B_s$. Peaking background are defined as three-body $B$ decays modes. The shapes and yields of the peaking backgrounds  are obtained from simulation. 
The invariant mass spectra of the  $\pi^\pm K^+  K^- $ and $\pi^\pm \pi^+  \pi^-$  candidates are shown in Fig.~\ref{MassFit}.

The signal yields obtained are $N(KK\pi)=1870\pm133$ and $N(\pi\pi\pi)=4904\pm 148$, and the raw asymmetries are $A_{raw}(K\!K\pi)=-0.143\pm 0.040$ and $A_{raw}(\pi\pi\pi) = 0.124\pm 0.021$, where the uncertainties are statistical. The CP asymmetries are expressed in terms of the 
measured raw asymmetries, corrected for effects induced by the detector acceptance and interactions of final-state pions with matter, as well as for a 
possible $B$-meson production asymmetry. To take care of these effects we use these informations from the control channel $ B^\pm \to J/\Psi K^+ $ with 
$J/\Psi \to \mu⁺ \mu^-$. 

The methods used in estimating the systematic uncertainties of the signal model, combinatorial background, peaking background, and acceptance correction are the same as those used in Ref.~\cite{LHCb-PAPER-2013-027}. 
For $\pi^\pm K^+ K^- $ decays, we also evaluate a systematic uncertainty due to the partially reconstructed background model by varying the mean and resolution according to the difference between simulation and data obtained from the signal component. The complete list of systematic and uncertainties  can be obtained in Ref. \cite{KKpiand3pi}. 

The results obtained for the inclusive CP asymmetries of the $ B^\pm \to \pi^\pm K^+  K^- $ and $ B^\pm \to \pi^\pm \pi^+  \pi^- $   decays are
\begin{eqnarray}
A_{CP}(B^\pm \to \pi^\pm K^+  K^-  )  \! &=& \!-0.141 \pm  0.040  \pm  0.018 \pm 0.007  ,  \nonumber  \\ [1mm]
A_{CP}( B^\pm \to \pi^\pm \pi^+  \pi^- )  \! & = &\!  0.117 \pm 0.021  \pm  0.009 \pm 0.007  ,  \nonumber
\end{eqnarray}
where the first uncertainty is statistical, the second is the experimental systematic, and the third is due to the CP  asymmetry of the control channel 
$ B^\pm \to J/\Psi K^+ $  reference  mode~\cite{PDG2012}.

\section{ CP asymmetry distribution in Dalitz phase space} 
\begin{figure*}[htb]
\centering
\includegraphics[width=0.49\linewidth]{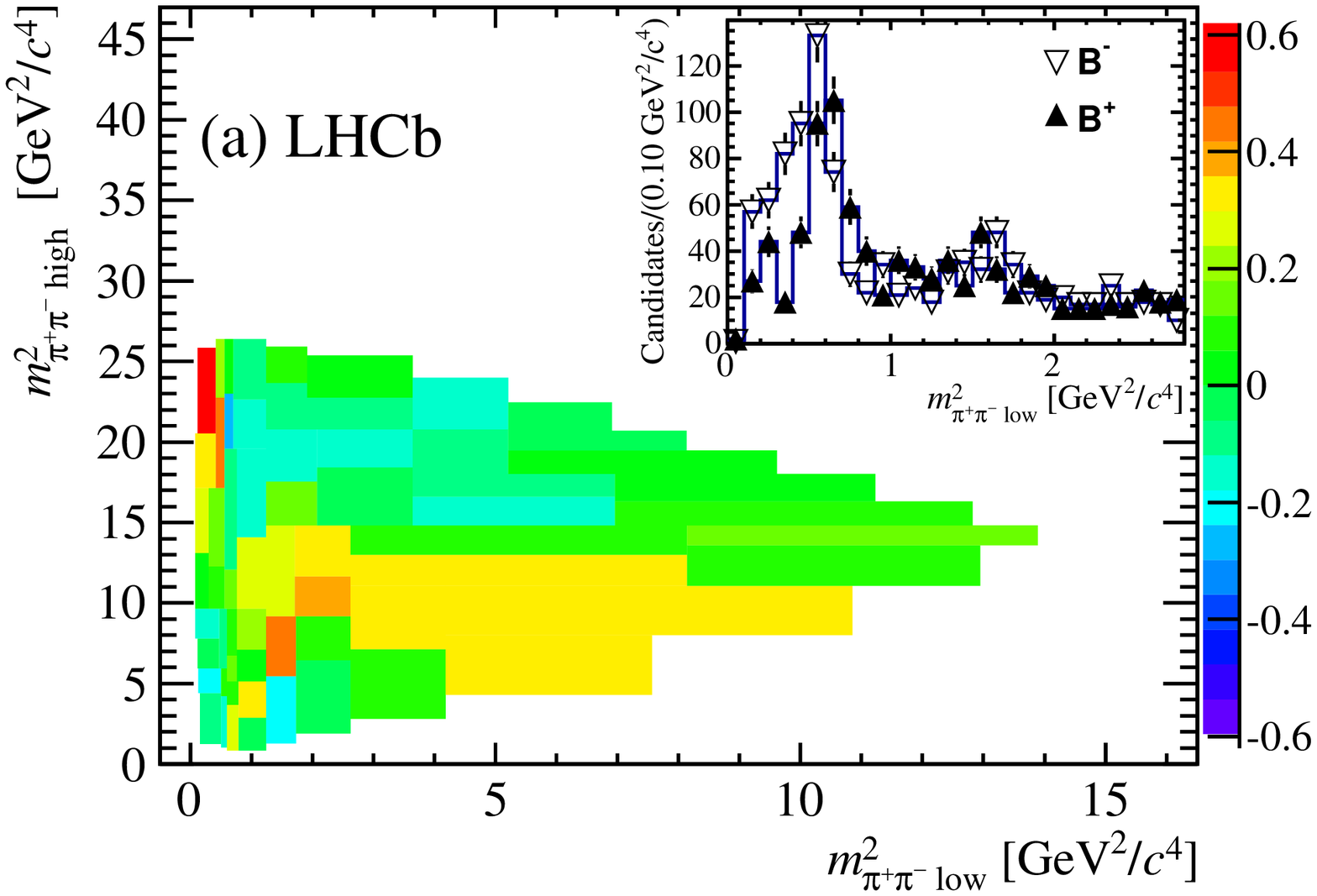}
\includegraphics[width=0.49\linewidth]{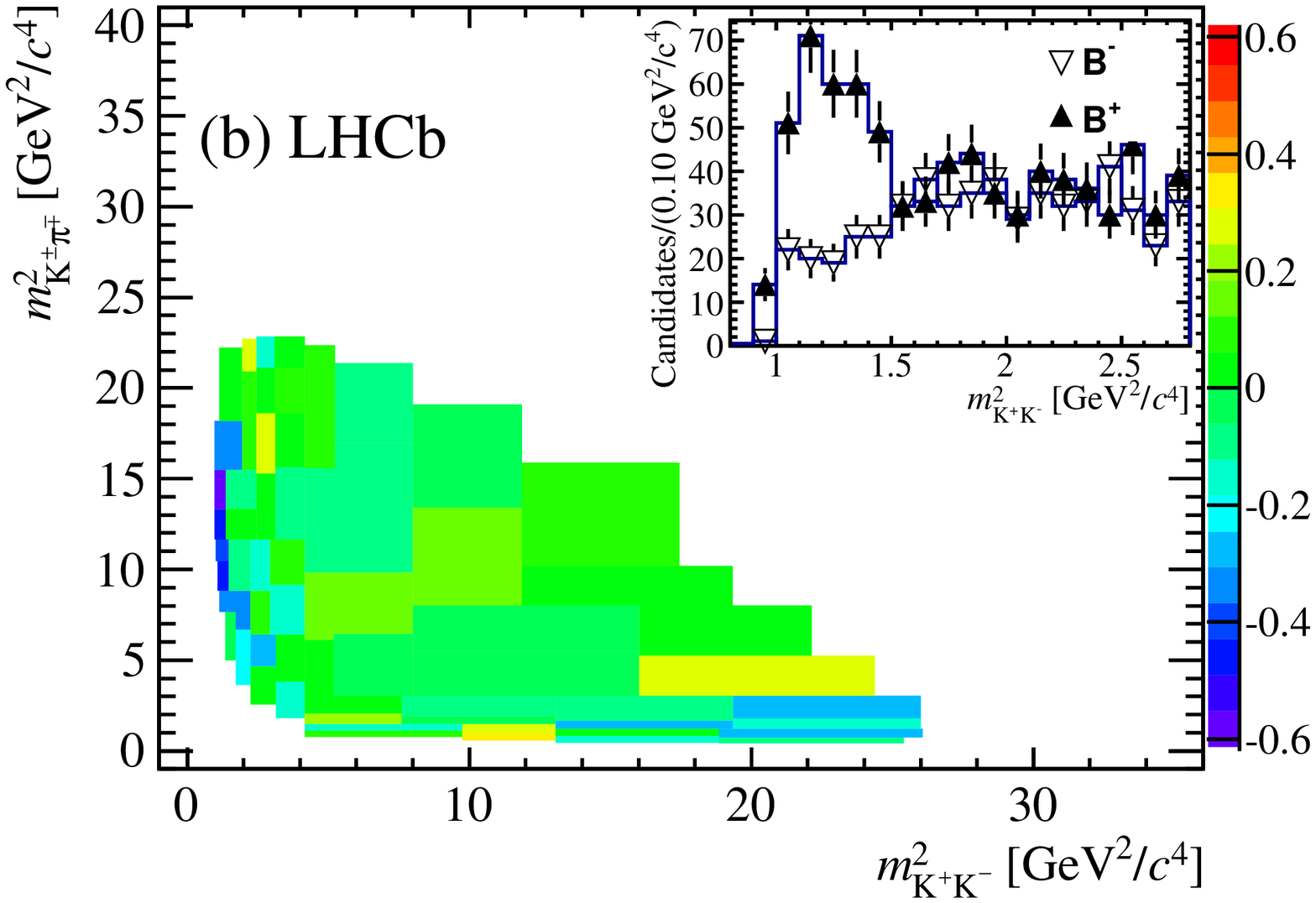}
\caption{
Asymmetries of the number of events (including signal and background) in bins of the Dalitz plot,  for (a) $ B^\pm \to \pi^\pm \pi^+  \pi^- $  and  (b)~
  $ B^\pm \to \pi^\pm K^+  K^- $ decays.  
The inset figures show the projections of the number of events in bins of  (a) the  variable $M^2_{high}(\pi\pi) > 15  $  and (b) the  variable $M^2(KK)$. 
The distributions are not corrected for efficiency.
}
\label{Mirandizing}
\end{figure*}

In addition to the inclusive charge asymmetries, we study the asymmetry distributions in the two-dimensional phase space of two-body invariant masses \cite{KKpiand3pi}. 
The Dalitz plot distributions in the signal region are divided into bins with approximately equal numbers of events in the combined $B^-$  and $B^+$ samples. 
Figure~\ref{Mirandizing} shows the raw asymmetries (not corrected for efficiency),  computed using the number of negative ($N^{-}$) and positive ($N^{+}$) entries in each bin of the $ B^\pm \to \pi^\pm K^+  K^- $ and $ B^\pm \to \pi^\pm \pi^+  \pi^- $ Dalitz plots.
The $ B^\pm \to \pi^\pm \pi^+  \pi^- $ Dalitz plot is symmetrized.

\begin{figure*}[htb]
\centering
\includegraphics[width=0.48\linewidth]{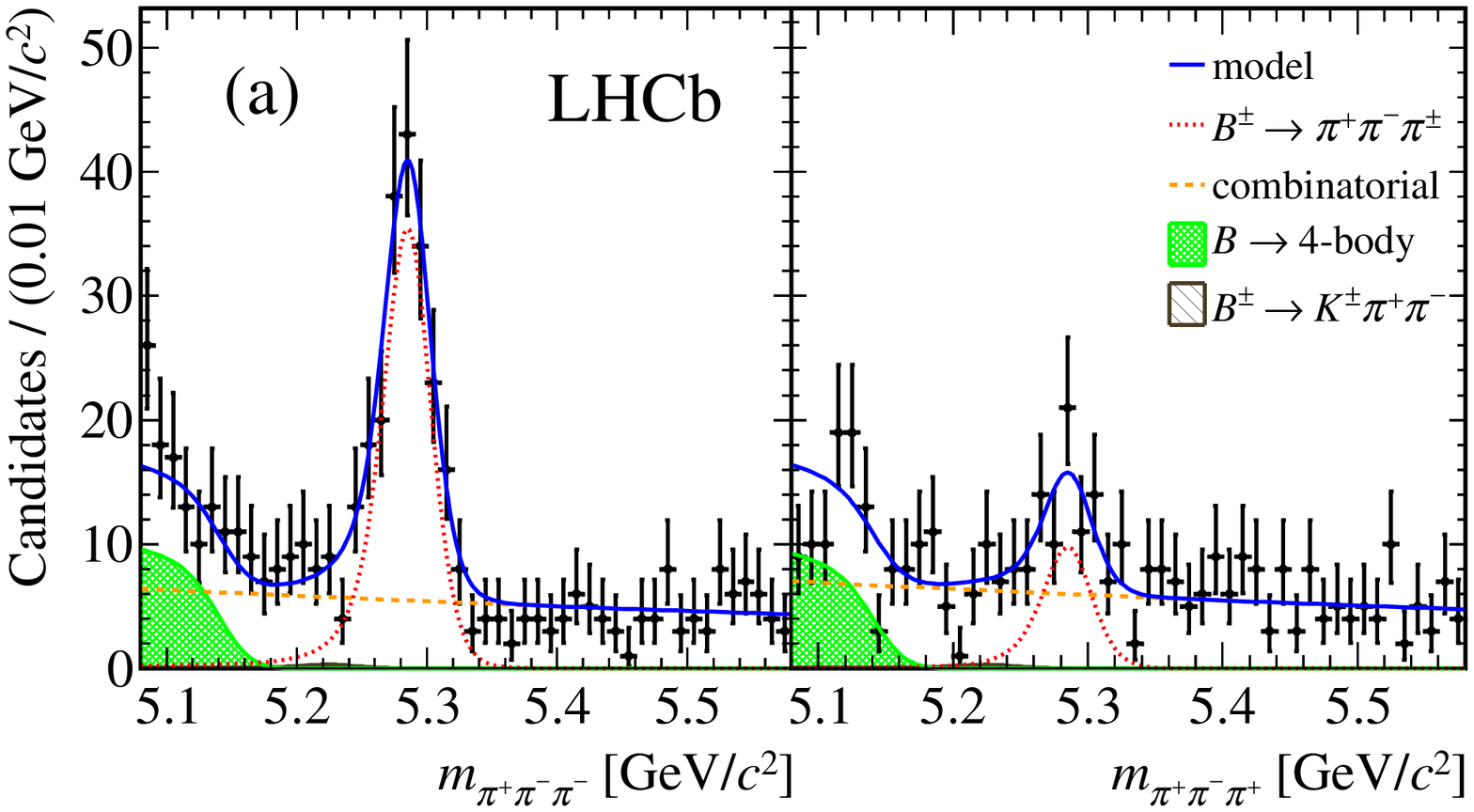}
\hspace{0.2cm}
\includegraphics[width=0.48\linewidth]{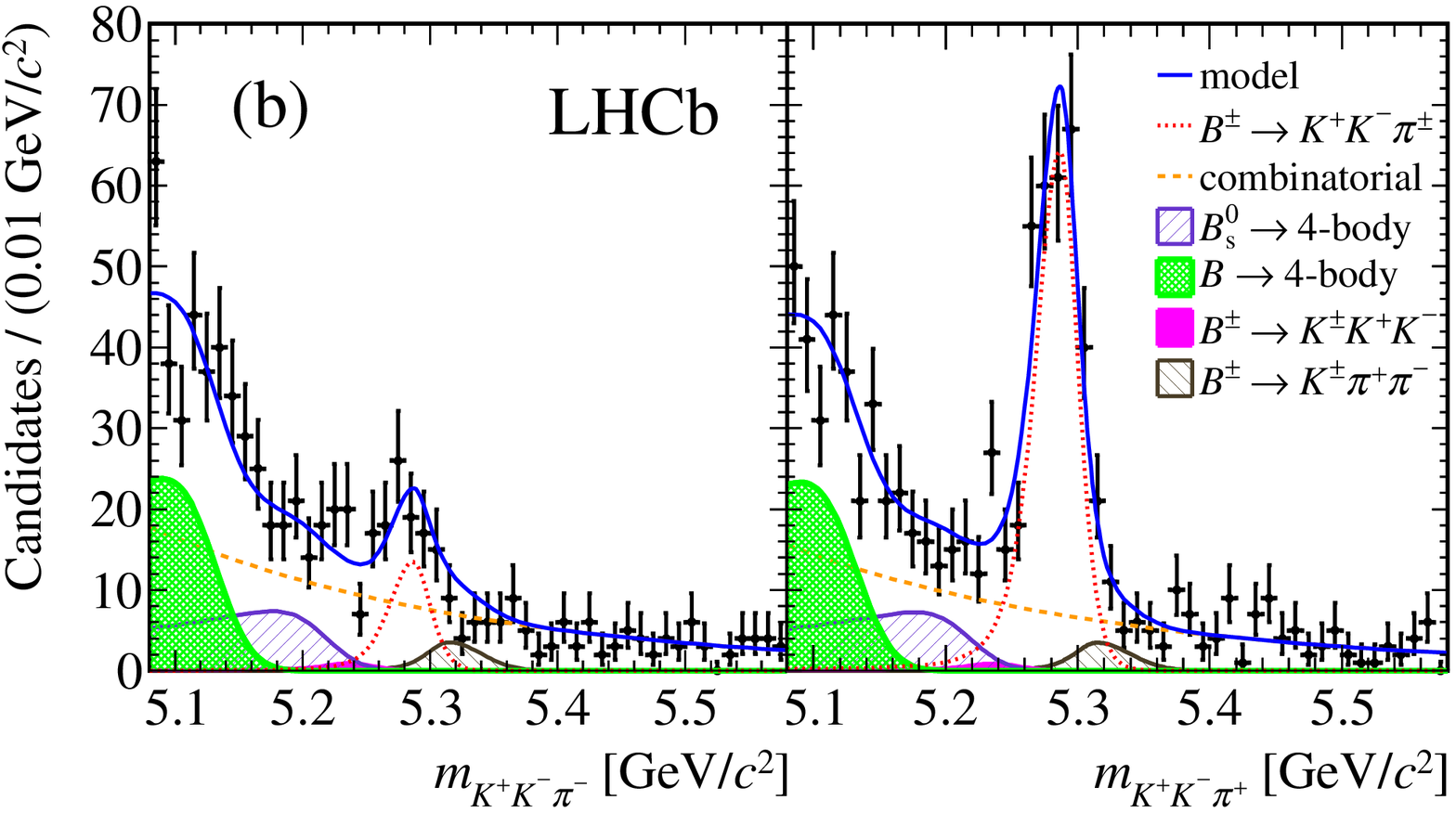}
\caption{Invariant mass spectra of (a) $ B^\pm \to \pi^\pm \pi^+  \pi^- $ decays in the region  $M^2_{high}(\pi\pi) > 15  $ and $M^2_{low}(\pi\pi) < 0.4 {\rm ~ GeV}^2 $   , and  (b)  $ B^\pm \to \pi^\pm K^+  K^- $  decays in the $M^2(KK) < 1.5 {\rm ~ GeV}^2$
region . The left panel in each figure shows the $B^-$  modes and the right panel shows the $B^+$ modes. The results of the unbinned maximum 
likelihood fits are overlaid. 
}
\label{MassFitRegion}
\end{figure*}

The CP asymmetries are further studied in the regions where large raw asymmetries are found. 
The regions are defined as $M^2_{high}(\pi\pi) > 15  $ and $M^2_{low}(\pi\pi) < 0.4 {\rm ~ GeV}^2 $ for the  $ B^\pm \to \pi^\pm \pi^+  \pi^- $  mode, and 
$M^2(KK) < 1.5 {\rm ~ GeV}^2$ for the  $ B^\pm \to \pi^\pm K^+  K^- $ mode.
Unbinned extended maximum likelihood fits are performed to the mass spectra of the candidates in these regions, using the same models as for the global fits. 
The spectra are shown in Fig.~\ref{MassFitRegion}. 
The resulting signal yields and raw asymmetries for the two regions are ${N^{\mathrm {reg}}(K\!K\pi)=342\pm28}$ and 
${A_{raw}^{\mathrm {reg}}(K\!K\pi)=-0.658\pm0.070}$ for the $ B^\pm \to \pi^\pm K^+  K^- $  mode, and ${N^{\mathrm {reg}}(\pi\pi\pi)=229\pm20}$ and 
${A_{raw}^{\mathrm {reg}}(\pi\pi\pi)=0.555\pm0.082}$ for the $ B^\pm \to \pi^\pm \pi^+  \pi^- $  mode. The CP asymmetries are obtained from the raw asymmetries  applying an acceptance correction, like above to the inclusive measurement.  The final results and the systematic uncertainties are produced  in a similar way that the ones of the inclusive CP asymmetries and are presented in Ref. \cite{KKpiand3pi}

These charge asymmetries are not uniformly distributed in the phase space. 
For  $ B^\pm \to \pi^\pm K^+  K^- $  decays, where no significant resonant contribution is expected, we observe  a very large negative asymmetry  concentrated in a restricted region of the phase space in 
the  low $K^+K^-$ invariant mass. 
For $ B^\pm \to \pi^\pm \pi^+  \pi^- $ decays, a large positive asymmetry is measured in the low         $M^2_{low}(\pi\pi) $ and high $M^2_{high}(\pi\pi) $ phase-space region, not clearly associated to a resonant state. The evidence presented here for  CP violation in  $ B^\pm \to \pi^\pm K^+  K^- $  and $B^\pm \to \pi^\pm \pi^+  \pi^- $   decays, along with the recent evidence for CP violation in $B^\pm \to K^\pm \pi^+  \pi^- $ and $ B^\pm \to K^\pm K^+  K^- $ decays~\cite{LHCb-PAPER-2013-027} and recent theoretical developments~\cite{Bhattacharya:2013cvn,IgnacioCPT,Xu:2013dta,PhysRevD.87.076007}, indicate new mechanisms for CP asymmetries, which should be incorporated in models for future amplitude analyses of charmless three-body $B$ decays.

\end{document}